\documentclass[twocolumn]{article}
\makeatletter
\@addtoreset{equation}{section}
\renewcommand\theequation{\thesection.\@arabic\c@equation}
\makeatother
\usepackage{amsmath,amssymb,latexsym}

\def\tit#1{``#1,''}
\def\ox{\otimes}

\newcommand{\ket}[1]{| #1 \rangle}
\newcommand{\bra}[1]{\langle #1 |}
\newcommand{\ktbr}[2]{\ket{#1}\bra{#2}}
\newcommand{\brkt}[2]{\langle #1 | #2 \rangle}
\newcommand{\vxv}[1]{\ket{#1}\bra{#1}}
\newcommand\Tr{\mathop\mathrm{Tr}}
\newcommand\IM{\mathop\mathrm{Im}}
\newcommand\RE{\mathop\mathrm{Re}}
\newcommand\Hil{\mathcal H}

\newcommand\R{\mathbb R}
\newcommand\Sv{\mathsf S}
\newcommand\Id{\mathbf 1}
\newcommand\Lin{\mathcal L}
\newcommand\Herm{\mathfrak H}
\newcommand\norm[1]{\left|{#1}\right|}
\newcommand\eq[1]{Eq.~(\ref{#1})}
\newcommand{\theor}[1]{\textit{Theorem}~\ref{#1}}
\newcommand{\lemma}[1]{\textit{Lemma}~\ref{#1}}
\newcommand\MEQ[1]{\mbox{$#1$}}

\newcommand{\mat}[1]{\left(\begin{smallmatrix} #1 \end{smallmatrix}\right)}
\newcommand{\matnorm}[1]{\langle\!\langle #1 \rangle\!\rangle}
\newcommand{\op}[1]{\hat{#1}}
\newcommand{\ems}{\sl\sffamily }
\def\eg{\textit{e.g.}, }
\def\Eg{\textit{E.g.}, }
\def\ie{\textit{i.e.}, }
\def\vz{\textit{viz}, }

\newtheorem{theo}{Theorem}[section]
\newtheorem{lem}{Lemma}[section]

\title{\bf\Large On Symmetric Sets of Projectors\\
 \large for Reconstruction of a Density Matrix}
\date{}
\author{Alexander Yu.\ Vlasov}
\begin{document}
\maketitle
\begin{abstract}
 In this work are presented sets of projectors for reconstruction of a
density matrix for an arbitrary mixed state of a quantum system with the
finite-dimensional Hilbert space.  It was discussed earlier \cite{tens} a
construction with $(2n-1)n$ projectors for the dimension $n$. For $n=2$ it is a
set with six projectors associated with eigenvectors of three Pauli matrices,
but for $n>2$ the construction produces not such a `regular' set. In this paper
are revisited some results of previous work \cite{tens} and discussed another,
more symmetric construction with the Weyl matrix pair (as the generalization
of Pauli matrices). In the particular case of prime $n$ it is the {\em
mutually unbiased} set with $(n+1)n$ projectors. In appendix is shown an
example of application of complete sets for discussions about separability and
random robustness.
\end{abstract}
\sloppy

\section{Introduction}

Let us consider \cite{tens} the $n$-dimensional Hilbert space $\Hil_n$ and
a density matrix \MEQ{\op\rho\in\Hil_n^*\ox\Hil_n} of a quantum system. Let we
have also a set of $N$ elements $\ket{v_\alpha} \in \Hil_n$ and the set
$\Sv_N(\Hil_n)$ of projectors $\op P_\alpha=\op P(v_\alpha)=\vxv{v_\alpha}$.
Each projector here may describe a probability of an outcome of some measurement
\begin{equation}
p_\alpha = \Tr(\op P_\alpha\op \rho), \quad \alpha = 1, \ldots, N
\label{prob}
\end{equation}
and so $N$ projectors via \eq{prob} produce some formal linear map
\begin{equation}
 \op\Lin_\Sv : \Herm(n) \to \R^N
\label{lins}
\end{equation}
from $n^2$-dimensional {\em real} space\footnote{A Hermitian matrix $H$
may be described by $n^2$ real parameters: $n$ real elements $H_{kk}$
and $n(n-1)/2$ pairs of elements $\{\RE(H_{kl}), \IM(H_{kl}); k>l\}$.}
$\Herm(n)$ of $n \times n$ Hermitian matrices $\op\rho$ to the formal vector
of probabilities $p_\alpha$ defined by \eq{prob}.

It shold be mentioned, that a physical density matrix also must have the
trace {\em one} and to be non-negative definite, but because the space of such
matrices is not linear, it is useful to introduce some constructions
for the linear space of Hermitian matrices, \eg it is clear, that
existence of an inverse map to $\op\Lin_\Sv$ is the sufficient condition for
reconstruction of the density matrix.

\section{Preliminaries}

\subsection{Classification}

Let us recollect a classification of sets $\Sv_N(\Hil_n)$ suggested in
\cite{tens}:
\begin{itemize}
 \item {\bf Representative:} exists {\em right inverse} of $\op\Lin_\Sv$,
 \ie it is possible to reconstruct the Hermitian matrix using the formal vector
 of probabilities $\R^N$.
 \item {\bf Minimal representative:} exists `usual' {\em inverse} of
 $\op\Lin_\Sv$,  \ie the representative set with $N=n^2$ elements and an
 isomorphism $\op\Lin^{-1}_\Sv: \R^{n^2} \to \Herm(n)$.
 \item {\bf Affine minimal:} it is possible to use the set with $N=n^2-1$
 and the condition $\Tr(\op\rho)=1$ to reconstruct a density matrix, \eg
 it is the minimal representative set without one vector.
 \item {\bf Complete:} $\Sv_N(\Hil_n)$ is representative and may be
 constructed by using a union of few orthogonal bases in $\Hil_n$.
 \item {\bf Almost perfect:} $\Sv_{mn}(\Hil_n)$ is the complete set constructed by
 using a {\em disjoint} union of $m$ orthogonal bases in $\Hil_n$, \ie $N=mn$.
 \item {\bf Perfect:} the set is almost perfect and each element is orthogonal
 {\em only} to $n-1$ vectors from its own basis.
\end{itemize}
It is useful to introduce in the present paper two new kinds of complete sets:
\begin{itemize}
 \item {\bf Mutually unbiased:} the perfect set constructed as a union of
 {\em mutually unbiased bases} (MUB) \cite{mub}, \vz for any two elements
 of {\em different} bases we have $\norm{\brkt vw}^2=1/n$.
 \item {\bf Symmetric:} the almost perfect set with the possibility of a
 transformation between any two bases using symmetries of $\Sv_{mn}(\Hil_n)$,
 \ie the unitary automorphisms of the corresponding set of vectors
 $\ket{v_\alpha}$,  $\alpha = 1,\ldots,mn$.
\end{itemize}

\subsection{Some previous results}

Let us mention few other facts proved in \cite{tens}:
\begin{theo} \label{th123}
Three properties are equivalent:
\begin{enumerate}
 \item\label{R1}
  A set of projectors $\Sv_N(\Hil_n)$ is representative.
 \item\label{R2}
  Any complex $n\times n$ matrix may be represented as a linear combination
  of the projectors $\op P(v_\alpha)$, \MEQ{\alpha=1,\ldots,N} with
  {\ems complex} coefficients.
 \item\label{R3}
  Any Hermitian $n\times n$ matrix may be represented as a linear combination
  of the projectors $\op P(v_\alpha)$, \MEQ{\alpha=1,\ldots,N} with
  {\ems real} coefficients.
\end{enumerate}
\end{theo}

One consequence of the \theor{th123} is \cite{tens}:
\begin{lem} \label{minset}
An (almost) perfect set may not be minimal, becasue it must have at least
$N=n^2+n$ elements.
\end{lem}
\noindent{\bf Proof:} Due to the \theor{th123} we must have the dimension of the
linear span of $\op P_\alpha$ not less than $n^2$. For the (almost) perfect set
with $m$ different bases the dimension is also not bigger than $mn-m+1$ due
to $m$ different presentation of same element, the unit matrix, as sum
of all $n$ projectors for a basis. So $m \ge n+1$ (\ie $mn-m+1 \ge n^2$)
and $N = n^2+n$ is the lower limit for a perfect set.  $\Box$

\noindent{\em Note:}
This limit is satisfied for mutually unbiased sets discussed below.

The complex decomposition mentioned in the \theor{th123} let us use a
convenient non-Hermitian basis like the set of $n^2$ matrices $\op E^{(kl)}$
with one unit in a cell $(k,l)$, \ie $(\op E^{(kl)})_{ij}=\delta_{ki}\delta_{lj}$
or, less formally, $\op E^{(kl)}=\ktbr kl$. Decomposition in such a basis is
simpler, but due to the \theor{th123} may be used in proofs of representativity
as well.
\Eg the coefficients of the decomposition may be simply found, if to
introduce some scalar product on the linear space of the matrices, say
\begin{equation}
 \matnorm{\op A , \op B}_* \equiv \Tr(\op A\,\op B^*).
\label{ortmat}
\end{equation}

This product is also in agreement with \eq{prob},
$p_\alpha=\matnorm{\op P_\alpha,\op\rho}_*$. Of course, it is always possible to
produce a Hermitian basis with the property
$\matnorm{\op H_\alpha,\op H_\beta}_*=\delta_{\alpha\beta}$ by application of
the standard Gram-Schmidt procedure, but the simpler non-Hermitian basis
$\op E^{(kl)}$ already has the property of orthogonality\footnote{Here $(kl)$
may be considered as a multi-index, or `linearized' as $\alpha=(k-1)\,n+l$.}
in the matrix norm \eq{ortmat}
$$ \matnorm{\op E^{(kl)}\!\!,\op E^{(ij)}}_* = \Tr(\ktbr kl\,\ktbr ji)
 =\brkt lj \,\brkt ik = \delta_{ki}\delta_{lj}.
$$

The $\op E^{(kl)}$ basis has also other useful property: tensor products of
such matrices from bases in dimensions $n$ and $l$ produce a basis in
dimension $nl$ with $(nl)^2$ matrices of same kind (\vz with only
one nonzero element, the unit in some cell). Such a property is usefull for
other important theorem \cite{tens}:

\begin{theo}[Composition theorem] \label{thcomp}
{\ }\\
Let $\Sv_N(\Hil_n)$ and $\Sv_P(\Hil_p)$ are {\ems representative (complete,
almost perfect)} sets with $N$ and $P$ vectors for two Hilbert spaces with
dimensions $n$ and $p$. Then a set $\Sv_{NP}(\Hil_n\ox\Hil_p)$ based on $NP$
tensor products of elements from both sets is, respectively, {\ems
representative (complete, almost perfect)} set for the composite system, but
such a tensor product of {\ems perfect} sets is only {\ems almost perfect}.
\end{theo}

\noindent{\em Note:}
It should be added, that a tensor product of {\em symmetric} sets is also
{\em symmetric}, because a tensor product of two symmetries, \ie linear
transformations between two bases, is again the symmetry, but a tensor product
of {\em mutually unbiased} sets is {\em not mutually unbiased}, because there
are orthogonal vectors in different bases.

\subsection{Examples of the sets}

Examples of representative and complete sets were presented in \cite{tens}.
Let us choose a basis $\ket{k}$, $k=1,\ldots,n$ of the Hilbert space,
then a representative set with $n^2$ projectors may be constructed
using $n+n(n-1)/2+n(n-1)/2=n^2$ vectors presented below \cite{tens,fin}:
\begin{center}
\begin{tabular}{|l|l|}
\hline
$n$ vectors of the basis:& $\ket{k}$, \\
$\frac{n(n-1)}{2}$ vectors:&
$\frac{1}{\sqrt2}(\ket{k}+\ket{l})$, $k < l$,\\
$\frac{n(n-1)}{2}$ vectors:&
$\frac{1}{\sqrt2}(\ket{k}+i\ket{l})$, $k < l$.\\
\hline
\end{tabular}
\end{center}

The representative set is not complete, but it is possible to use
yet another $n(n-1)$ vectors: to produce a complete set with $(2n-1)n$
projectors
\begin{center}
\begin{tabular}{|l|l|}
\hline
$\frac{n(n-1)}{2}$ vectors:&
$\frac{1}{\sqrt2}(\ket{k}-\ket{l})$, $k < l$,\\
$\frac{n(n-1)}{2}$ vectors:&
$\frac{1}{\sqrt2}(\ket{k}-i\ket{l})$, $k < l$.\\
\hline
\end{tabular}
\end{center}

The set is complete, because it is the union of \MEQ{n^2-n+1} bases:
\begin{itemize}
\item An initial basis: $\ket{k},\ k=1,\ldots,n$.
\item $n(n-1)/2$ bases produced by substitution of two elements in
initial basis: $\frac{1}{\sqrt2}(\ket{k}+\ket{l})$ and
$\frac{1}{\sqrt2}(\ket{k}-\ket{l})$ instead of $\ket{k}$ and $\ket{l}$.
\item $n(n-1)/2$ bases produced by substitution:
$\frac{1}{\sqrt2}(\ket{k}+i\ket{l})$ and
$\frac{1}{\sqrt2}(\ket{k}-i\ket{l})$ instead of $\ket{k}$ and $\ket{l}$.
\end{itemize}
Certainly, this complete set is not a disjoint union for $n>2$, because
here exist different bases with $n-2$ common elements.

\section{Symmetric complete sets of projectors}

The example above contains $(2n-1)n$ elements, but only for $n=2$ it is a
{\em perfect} and {\em symmetric} set.  In such a case there are three bases
with two vectors corresponding eigenvectors of three Pauli matrices
$$
 \op\sigma_x = \mat{0&1\\1&0},~
 \op\sigma_y = \mat{0&-i\\ i&0},~
 \op\sigma_z = \mat{1&0\\0&-1}.
$$
Similar aproach may be used in a highter dimension due to
auxiliary lemmas sugested below.

\subsection{Auxiliary lemmas}

Let us consider an unitary matrix $\op M$ and the orthonormal basis produced
by eigenvectors $\ket{\mu_k}$, $k=1,\ldots,n$
\begin{equation}
 \op M\, \ket{\mu_k} = \lambda_k \ket{\mu_k},
\quad \brkt{\mu_k}{\mu_j} = \delta_{ij}.
\label{Opsi}
\end{equation}

Orthogonal projectors associated with such a matrix are defined as
\begin{equation}
 \op P_{\op M,k} \equiv \vxv{\mu_k}.
\label{POk}
\end{equation}

\begin{lem} \label{lemOP}
Any power $\op M^p$ of the matrix \eq{Opsi} may be expressed as a linear
combination of associated projectors $\op P_{\op M,k}$ \eq{POk}.
\end{lem}

\noindent{\bf Proof:}
$$
 \op M^p = \op M^p\!\underbrace{\sum_{k=1}^n \ktbr{\mu_k}{\mu_k}}_{\Id}
 {=} \sum_{k=1}^n \lambda_k^p \ktbr{\mu_k}{\mu_k}
  = \! \sum_{k=1}^n \lambda_k^p \op P_{\op M,k}
$$
\vskip -5ex
\hfill$\Box$
\vskip 3ex

\begin{lem} \label{lemain}
Let we have some set of unitary matrices $\op M_{(i)}$ and any complex matrix may be
expressed as a sum of powers $\op M_{(i)}^p$ with complex coefficients\footnote{%
The power zero, $\op M^0=\Id$ is also taken into account here.}, then it is
possible to use orthonormal bases of eigenvectors of $\op M_{(i)}$ for
construction of a {\ems complete} set of projectors.
\end{lem}
{\bf Proof:} Due to the \lemma{lemOP} we have the decomposition of any complex
matrix using eigenvectors, due to the \theor{th123} the set is representative
and it is the union of orthogonal bases, \ie complete.
$\Box$

\Eg it is true for Pauli matrices because together with the unit they
are basis of complex $2 \times 2$ matrices.

\subsection{Weyl pair}

An analogue of such unitary basis in a higher dimension $n$ \cite{QEC} is $n^2$
matrices $\op U^k \op V^l$, $k,l = 0,\ldots,n-1$, where $\op U,\op V$ is
the Weyl pair
\cite{WeylGQM}
\begin{equation}
 \op U = \mat{0&1&0&\ldots&0\\0&0&1&\ldots&0\\
 \vdots&\vdots&\vdots&\ddots&\vdots\\0&0&0&\ldots&1\\1&0&0&\ldots&0}\!,
 ~~
 \op V = \mat{1&0&0&\ldots&0\\0&\zeta&0&\ldots&0 \\
 0&0&\zeta^2&\ldots&0 \\ \vdots&\vdots&\vdots&\ddots&\vdots\\
 0&0&0&\ldots&\zeta^{n-1}}\!.
\label{WeylPair}
\end{equation}
(where $\zeta = \exp{2\pi i/n}$),
but the \lemma{lemain} does not use products of powers. It is necessary to
choose a minimal set of $\op U^a \op V^b$, $0 \le a,b < n$ with same eigenvectors
as the whole set.

\begin{lem}{\bf(Discrete version of von Neumann uniqueness theorem)}
\label{UjV}
Let $\op A,\op B \in SU(n)$,
\begin{equation}
 \op A \op B = \zeta^j \op B \op A,\quad \gcd(j,n)=1,
\label{UjVAB}
\end{equation}
(\vz $j$ and $n$ are coprime)
then exists a unitary transformation $\op S$:
\begin{equation}
 \op S\op A\op S^{-1} = \op U^j,\quad \op S\op B\op S^{-1} = \op V.
\label{UjVS}
\end{equation}
\end{lem}

\noindent{\bf Proof:} Let us rewrite \eq{UjVAB} as
\begin{equation} \tag{\ref{UjVAB}$'$}
 \op A\op B\op A^{-1} = \zeta^j \op B,
\label{ABA}
\end{equation}
but $\op A\op B\op A^{-1}$ and $\op B$ have same eigenvalues and so due to
\eq{ABA} $\op B$ and $\zeta^j \op B$ have same eigenvalues, but it is
possible only for a set of $n$ numbers $\zeta^{j\,k}$, $k = 0,\ldots,n-1$. If
$\gcd(j,n)=1$, this set after some permutation corresponds to the set
$\zeta^{k}$, $k = 0,\ldots,n-1$. It is precisely eigenvalues of $\op V$ and
so diagonalization of $\op B$ by transition $\op S$ to the basis of
eigenvectors of $\op B$ is just $\op V=\op S\op B\op S^{-1}$. In this new
basis another matrix is $\op A'=\op S\op A\op S^{-1}$ and it is clear from
\eq{ABA}, that $\op A'$ is the cyclic shift of eigenvectors of $\op V$ with
`step' $j$, \ie $\op A'=\op U^j$ and so \eq{UjVS} holds. $\Box$

Two particular examples of $\gcd(j,n)=1$ are \mbox{\{$j=1$, $\forall n$\}}
(see \cite{WeylGQM} for more details with this particular example) and
\{$\forall j$, $n$ is prime\} (it is a main application in this article).

It should be mentioned, that if $\gcd(j,n)=k$, $n=kl$, then matrices
are reducible, with $k$-dimensional subspaces corresponding to
equal eigenvalues, \eg it is tensor products like $\op V_l \ox \Id_k$.

\subsection{Constructions of symmetric sets}
\subsubsection{Prime dimension}

\begin{theo} \label{thprime}
 Let $n$ is a prime number, then eigenvectors of $n+1$ matrices:
 $\op U$, $\op U^m \op V$, $m = 0,\ldots,n-1$ produce
 a complete set of projectors.
\end{theo}

\noindent{\bf Proof:} 
 It can be written for $m=1,\ldots,n-1$:
$(\op U^m \op V)^l = \alpha (\op U^{ml \pmod{n}}\op V^l)$ with
complex $\alpha$ due to properties of the Weyl pair: $\op U\op V =\zeta\op V\op U$,
$\op U^n=\op V^n=\Id$. The equation $ml \pmod{n} = k$ for any $k,l$ always has
some solution $m$, if $n$ prime, because in such a case arithmetic
modulo $n$ is field. So any matrix $\op U^k \op V^l$, $k,l = 1,\ldots,n-1$
may be presented as a power of $\op U^m \op V$, and together with powers
of $\op U$ and $\op V$ it is any matrix $\op U^k \op V^l$, $k,l = 0,\ldots,n-1$,
but it is the basis. So any complex matrix may be expressed as a sum
of powers of $n+1$ matrices suggested above with complex coefficients
and due to the \lemma{lemain}, it is possible to use eigenvectors of the
matrices for construction of a complete set of projectors. $\Box$

\begin{theo}
 The complete set of projectors described above in the \theor{thprime}
 is also perfect, symmetric  and mutually unbiased.
\end{theo}
\noindent{\bf Proof:}
Any mutually unbiased set is perfect by definition. Let us prove,
that our set is mutually unbiased \cite{mubPhD} and symmetric. Any two
different matrices $\op A,\op B$ considered in the \theor{thprime} has property
$\op A\op B = \zeta^j \op B\op A$ for some $j=1,\ldots,n-1$ and the \lemma{UjV}
let us consider a new basis, there $\op A$ and $\op B$ may be rewritten as
$\op U^j$ and $\op V$. Elements of this new basis are eigenvectors $\ket{b_k}$
of $\op B$. In the basis $\op A$ is cyclic $j$-shift, and eigenvectors of $\op A$
may be written as $\ket{a_k} = \sum_{l=0}^{n-1}\zeta^{jkl}\ket{b_l}/\sqrt{n}$,
\ie $\norm{\brkt{a_k}{b_m}}=1/\sqrt{n}$, $\forall k,m$. It was
considered an arbitrary pair $\op A$, $\op B$ between $n+1$ matrices, \vz all bases
are mutually unbiased (see also \cite{mubPhD}).

Let us prove now, that the set is symmetric. If to show, that exists
a symmetry between any basis and eigenvectors of $\op V$ (\ie the initial basis
of the Hilbert space)\footnote{Of course unitary transformation between two basis
always exists, even $n!$  such transformations, but here is necessary to
find a symmetry of the set, \eg it must maps the set of all vectors to itself.}
then a symmetry of any two bases may be expressed via two such transformations
as $\op T_1 \op T_2^{-1}$. Due to the \lemma{UjV} (with $\op A=\op V$,
$\op B=\op U$) exists transformation $\op U \mapsto \op V$,
$\op V \to \op U^{-1}$ (it is discrete Fourier transform). It is the symmetry,
because maps $\op U^k\op V^l \mapsto \op V^k\op U^{-l}=\op V^k\op U^{n-l}$,
and so it is an automorphism for the set of operators $\op U^k\op V^l$ and
they eigenvectors. Due to the \lemma{UjV} (with $\op A=\op U$,
$\op B=\op V\op U^k$) exists a transformation $\op V\op U^k \mapsto \op V$,
$\op U \mapsto \op U$ and it is also the symmetry. Here are $n$ transformations
$\op T_j$ to the canonical basis of $\op V$, and $\op T_j\op T_k^{-1}$ is
the symmetry between two arbitrary bases. $\Box$.

\subsubsection{Non-prime dimension}

If the dimension is not prime, it is possible to use a tensor product of few
symmetric sets to construct a symmetric set in the composite dimension due to
the {\em Note} after the \theor{thcomp}. It was also mentioned, that a tensor
product of unbiased sets is not unbiased, but it is always almost perfect,
as a product of perfect sets. Really mutually unbiased bases exist not only
in prime dimensions, but for any power $p^l$ of prime \cite{mubPhD}.
The research of application of such bases for construction of a complete set
is an interesting problem, but it is outside of the scope of present work.

It should be mentioned, that $\op U^k\op V^l$ are an unitary basis in an
arbitrary dimension, and so bases of eigenvectors may be always used for
construction of a complete set due to the \lemma{lemain} (applied to the
`power' $p=1$). But only for prime dimension $n+1$ operators introduced in
the \theor{thprime} have all properties necessary for constructions used above.

If the dimension is not prime, there are following problems:
\begin{enumerate}
\item The whole set $\op U^k\op V^l$ may not be represented as powers of the
 $n+1$ matrices from the \theor{thprime}.
\item The operators $\op U,\op V$ have subspaces with equal eigenvalues and so
eigenvectors for such subspaces may be presented using arbitrary
combinations, \vz not in an unique way.
\item The \lemma{UjV} used for the proof of symmetry does not work if
$\gcd(j,n)>1$.
\end{enumerate}

So it is possible to construct a {\em complete} set of projectors using
products of the Weyl pair in any dimension, but it have less `regular' structure
if the dimension is not prime.

Using the tensor product structure with sets of prime dimensions $p_k$ for
$n=\prod_k p_k$, it is possible to construct almost perfect, symmetric sets
with $m=\prod_k(p_k+1)>n+1$, $N=mn$ elements, but at least for the power of
prime, exist mutually unbiased, \ie perfect sets with $m=n+1$ and the smallest size
$N$. Even if for products of different primes, mutually unbiased sets are not
exists, may be it is possible using eigenvectors of $\op U^k\op V^l$
at least construct complete, or (almost) perfect, or symmetric set with dimension
{\em smaller} than $\prod_k (p_k+1)$? Possibly the question devotes further
research.

\appendix
\section*{APPENDIX}
\section{Separability and robustness}\label{Rob}

To show application of a complete set of projectors, let us demonstrate a
proof for analogue of some theorem from \cite{SepVol,SepNMR} for an arbitrary
number of finite-dimensional quantum systems.

\begin{theo}[ZHSL-BCJLPS'] \label{thAff}
 A density matrix $\op\rho$ of any composite system may be represented as
\begin{equation}
 \op\rho = \alpha\op\rho_s - \beta \Id, \quad \alpha,\beta \in \R,~\alpha > 0,
\label{rho1}
\end{equation}
 where $\op\rho_s$ is a density matrix of a {\ems separable} state, \ie
\begin{equation}
 \op\rho_s = \sum_I \alpha_I \op\rho^{(1)}_I \ox \cdots \ox \op\rho^{(k)}_I,
 \quad \alpha_I > 0
\label{rhos}
\end{equation}
and each $\op\rho^{(i)}_I$ is a valid density matrix of $i$-th subsystem.
\end{theo}

\noindent{\bf Proof:} Let us consider a complete set of projectors for
each subsystem, then due to the \theor{thcomp} it is possible to construct
a set of projectors for the whole system as the tensor product. Using this
{\em representative} set, due to the \theor{th123}, it is possible to write
any density matrix as
\begin{equation}
 \op\rho = \sum_\alpha k_\alpha \op P_\alpha, \quad k_\alpha \in \R,
\label{rhoneg}
\end{equation}
where not all $k_\alpha$ are necessary positive, but instead of negative terms
it is possible to write
\MEQ{k_\alpha \op P_\alpha = k_\alpha \Id + (-k_\alpha)(\Id - \op P_\alpha)},
and for the {\em complete} set $\Id - \op P_\alpha$ always may be represented
as the sum of other projectors\footnote{Sum of all projectors for the given
basis is the unit and for a complete set each elements belongs to some basis.}
and so \eq{rhoneg} may be rewritten as
\begin{equation}
 \op\rho = \sum_\alpha k'_\alpha \op P_\alpha - k \Id, \quad k'_\alpha > 0,
\label{rho2}
\end{equation}
and because we use construction of the complete set as the tensor product
$\op P_\alpha = \op P_{\alpha_1} \ox \cdots \ox \op P_{\alpha_k}$,
\eq{rho2} coincides with \eq{rho1} after substitution \eq{rhos},
and $\op P_{\alpha_i}=\vxv{v_{\alpha_i}}$ is the valid density matrix
(of a pure state). $\Box$

This proof for composition of the arbitrary number of systems with the
arbitrary (possibly different) finite dimensions --- is a generalized analogue
of \cite{SepNMR} for two-dimensional systems and Pauli matrices. It should
be mentioned also, that a minimum of $\beta$ for different $\op\rho_s$ in
\eq{rho1} characterises a measure of separability of quantum systems
known as the random robustness \cite{Vidal}. The proof of the \theor{thAff}
above is constructive, but not necessary provides this minimal value of $\beta$.
Does it possible to suggest some optimization strategy using complete
sets of projectors? It is yet another interesting problem.


\def\bibpref{\ref{Rob}.}
\end{document}